\documentclass[pra,twocolumn,superscriptaddress,amsmath,amssymb]{revtex4-1}

\usepackage{graphicx}

\begin{document}

\title{Amplitude of waves in the Kelvin-wave cascade}

\author{V.\,B. Eltsov}
\affiliation{Department of Applied Physics, Aalto University, POB 15100, FI-00076 AALTO, Finland}

\author{V.\,S. L'vov}
\affiliation{Department of Chemical and Biological  Physics, Weizmann Institute of Science, Rehovot 76100, Israel}

\begin{abstract}
Development of experimental techniques to study superfluid dynamics, in particular, application of nanomechanical oscillators to drive vortex lines, enables potential observation of the Kelvin-wave cascade on quantized vortices. One of the first questions which then arises in analysis of the experimental results is the relation between the energy flux in the cascade and the amplitude of the Kelvin waves. We provide such relation based on the L'vov-Nazarenko picture of the cascade. Remarkably, the amplitude of the waves depends on the energy flux extermely weakly, as power one tenth. 
\end{abstract}

\def\km{k_{\rm min}}
\def\cLN{C_{_{\rm LN}}}
\def\rhos{\rho_{\rm s}}
\def\nus{\nu_{\rm s}}
\def\Ev{E_{\rm v}}
\def\kw{_{\rm kw}}
\def\LLkw{{\cal L}\kw}

\maketitle

In quantum turbulence, velocity fluctuations and vortex reconnections drive oscillating motion of quantized vortices -- Kelvin waves \cite{DonnellyBook}. Kelvin waves interact non-linearly and support a cascade of energy towards smaller length scales and larger wave numbers. In the currently accepted picture of quantum turbulence~\cite{Vinen2008}, a quasi-classical hydrodynamics energy cascade at length scales larger than the intervortex distance $\ell$ is followed, after some cross-over region, by the Kelvin-wave cascade at scales smaller than $\ell$ \cite{CascadesReview}. The mutual friction damps Kelvin waves very efficiently, and the cascade is expected to start to develop when the mutual friction $\alpha$ is well below $10^{-3}$ \cite{Boue2015}. As temperature and $\alpha$ decreases, the cascade extends to progressively smaller length scales and eventually, at the lowest temperatures, it is damped by emission of bosonic \cite{Vinen2001} or fermionic \cite{Silaev2012,Makinen2018} quasiparticles by the oscillating vortex cores.

The theory of the Kelvin-wave cascade was the subject of controversy \cite{KS,Vinen2005,LN1,LN2,Sonin}, until finally the L'vov-Nazarenko model got supported by numerical simulations \cite{Krstulovic2012,Baggaley2014}. The theory is built for a straight vortex with uniform occupation of Kelvin wave modes along the length. Such situation never occurs in a typical experiment on quantum turbulence. Recently, progress in experimental techniques \cite{LancNEMS,LeeMEMS,Kamppinen2019} enables controllable excitation of waves on straight or nearly straight vortices, see Fig.~1 for possible setups. Such experiments have potential to observe Kelvin-wave cascade directly and thus allow comparison to the theory. One of the first questions which analysis of such experiments poses is the relation of the energy flux carried by the cascade (observed, e.g., as an increase of the damping of a nanomechanical agitator) to the amplitude of the excited Kelvin waves. We provide such relation in this work.

We assume that the Kelvin-wave cascade on a vortex of length $L$ [cm] carries the energy flux $\tilde\epsilon$ [erg/s] and starts from the wave number $\km$ [cm$^{-1}$]. Our goal is to find the amplitude $A_k$ [cm] of the Kelvin wave with the wave number $k$ [cm$^{-1}$]. We start by noting that
in the local induction approximation the energy of a vortex line $\Ev$ is given by the product of its length $L$ and the vortex tension $\nus$
\begin{equation}
  \Ev = \nus L\,, \quad \nus = \rhos  \frac{\kappa^2 \Lambda}{4 \pi  }\, , \quad  \Lambda = \ln \Big ( \frac {\ell}{a_0} \Big )\ .
  \label{Ev}
\end{equation}
Here $\rhos$ is the superfluid density, $\kappa$ is the circulation quantum, $a_0$ is the vortex core radius and $\ell$ is the mean intervortex spacing or the size of the enclosing volume, in the case of a single vortex. For a spiral Kelvin wave of the radius $A_k$ and wavelength $\lambda_k = 2\pi/k$, the increase of the length compared to that of the straight vortex is
\begin{equation}
  L_k = \left( \sqrt{\lambda_k^2 + (2\pi A_k)^2} - \lambda_k\right)
  \frac L{\lambda_k} \approx
  L \,\frac{2\pi^2 A_k^2}{\lambda_k^2}\,,
  \label{Lk}
\end{equation}
where we assumed that $A_k \ll \lambda_k$. Thus the total energy due to Kelvin waves is
\begin{align}
\begin{split}
  E\kw &= \sum_{k=\pm\km}^{\pm\infty} \nus L_k
  = L \sum_{k=\km}^\infty \nus A_k^2 k^2\\
  &= L\, \frac\nus\km \int_{\km}^\infty A_k^2 k^2 \,dk\ .
  \label{EA}
\end{split}\end{align}
Comparing this result to the expression of the energy via the Kelvin-wave frequency $\omega_k$ and the combined occupation number $N_k$ for modes with $\pm k$ \cite{KW2}
\begin{equation}
  E\kw = \rhos L \int\limits _{\km}^\infty E_k\,dk\,,\ 
  E_k = \omega_k N_k\,,\ 
  \omega_k = \frac{\kappa\Lambda}{4\pi} k^2\,,
  \label{Ekw}
\end{equation}
we find
\begin{equation}
  A_k^2 = \frac\km\kappa N_k \ .
  \label{AkNk}
\end{equation}

The L'vov-Nazarenko  spectrum  is \cite{KW2}
\begin{subequations}\label{LN}
  \begin{align} \label{LNE}
E_k &= \cLN\frac{\kappa \Lambda \epsilon^{1/3}}{\Psi^{2/3}k^{5/3}}\,,\qquad
\cLN \approx 0.304\,,
\\ 
 \Psi&=  \frac{8\pi}{\Lambda \kappa^2}\int_{\km}^\infty E_k dk\ .
\label{LNPsi}
 \end{align}\end{subequations}
Here $\epsilon$ is the energy flux per unit length and per unit mass. It is related to the flux $\tilde\epsilon$ as
\begin{equation}
\epsilon = \frac{\tilde\epsilon}{L \rhos}\,,
\qquad [\epsilon] = \frac{\mathrm{cm}^4}{\mathrm{s}^3}\ .
\label{eps}
\end{equation}
Solving Eq.~(\ref{LN}) for $\Psi$ we get
\begin{equation}
\Psi = \frac{(12\pi\cLN)^{3/5} \epsilon^{1/5}}{
\kappa^{3/5} \km^{2/5}}
\label{psisol}
\end{equation}
and from Eq.~(\ref{AkNk}) finally
\begin{align}\begin{split}
\label{Asol}
A_k^2 &= 2\left(\frac{2 \pi^3 \cLN^3}{9}\right)^{1/5}
\frac{\km^{19/15}\epsilon^{1/5}}{\kappa^{3/5} k^{11/3}} \\  
&\approx 1.4 \, \frac{\km^{19/15}}{\kappa^{3/5} k^{11/3}} 
\left( \frac{\tilde\epsilon}{L \rhos} \right)^{1/5}\ .
\end{split}\end{align}
Checking dimensions we find correctly $[A_k^2] = \textrm{cm}^2$. Note that $A_k \propto \tilde\epsilon^{1/10}$. Thus determination of the amplitude from the energy flux should be relatively reliable, while the reverse procedure is bound to be very uncertain.

\begin{figure}[t]
\centerline{\includegraphics[width=0.8\linewidth]{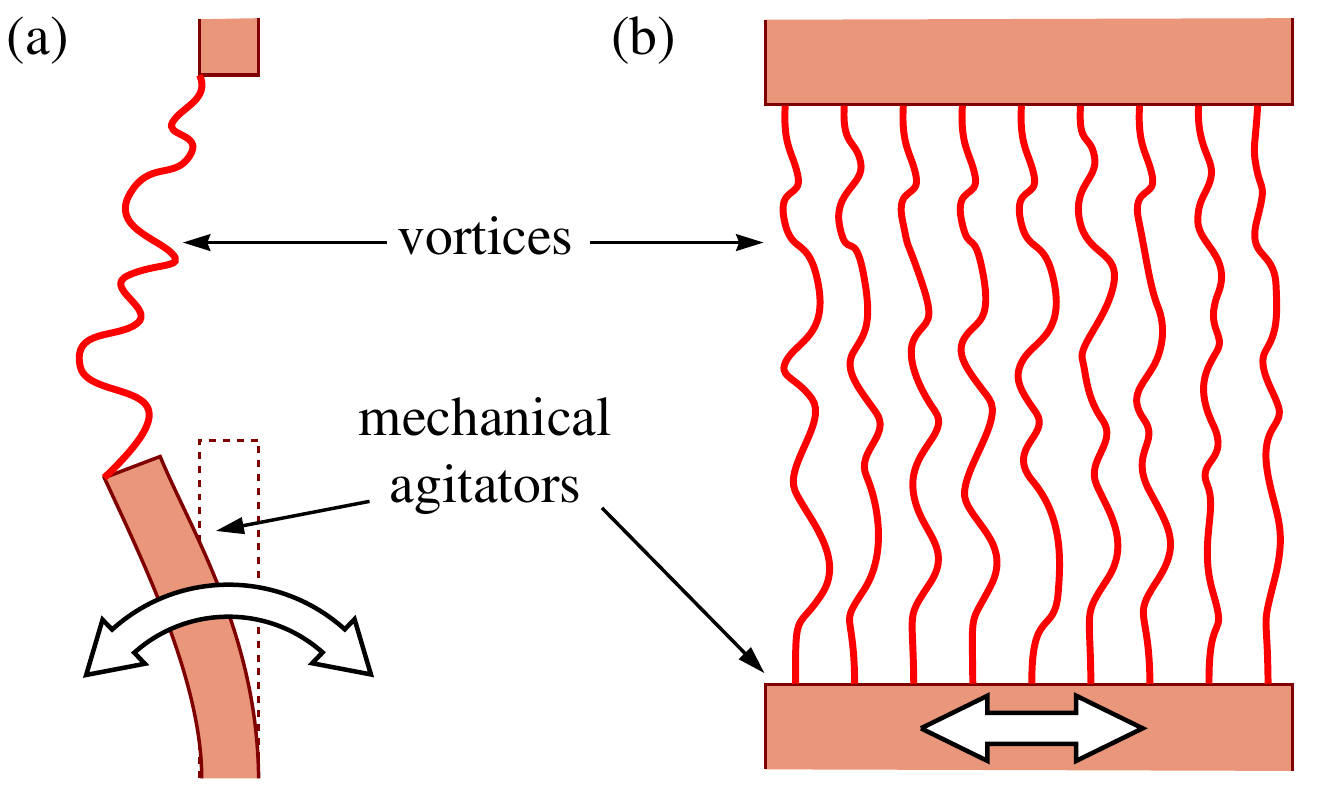}}
\caption{Fig.~1. Example configurations of vortex lines, agitated to generate Kelvin waves. (a) A single vortex, attached to an oscillating device. (b) An array of vortices, stretched between parallel plates and agitated by shear or torsional oscillations of the plates.}
\end{figure}

The total increase of the vortex line length due to Kelvin waves can be found from the energy as $L\kw = E\kw/\nus$, where $E\kw$ is given by Eqs.~(\ref{Ekw}), (\ref{LNE}) and (\ref{psisol}):
\begin{equation}
L\kw = \frac{E\kw}\nus = L \,\frac{2^{1/5}(3\pi \cLN)^{3/5} \epsilon^{1/5}}
{\kappa^{3/5}\km^{2/5}} \ . 
\end{equation}
Thus for the relative increase we get a simple formula
\begin{equation}
\frac{L\kw}{L} = \frac{E\kw}{E_{\rm v}} = \frac\Psi2 \ .
\label{LkwL}
\end{equation}

In cases, where instead of a single vortex, one considers a vortex array with the total length $L$ occupying volume $V$ with the density ${\cal L} = L/V = \ell^{-2}$ (Fig.~1b), it might be more convenient to operate with the standard 3-dimensional energy flux $\varepsilon$ per unit mass and unit volume, $[\varepsilon] =$cm$^2$s$^{-3}$. Having geometry of Fig.~1b in mind, it is easy to see that $\varepsilon = \epsilon {\cal L}$. Then for the increase $\LLkw$ of the vortex-line density due to Kelvin waves, we find using Eqs.~(\ref{psisol}) and (\ref{LkwL})
\begin{align}\begin{split}
\frac\LLkw{\cal L} = \frac\Psi2 
&=\Big[
\frac{2(3\pi\cLN)^3 \varepsilon}
{b^2 {\cal L}^2 \kappa^3} \Big]^{1/5}\\
&\approx 2.2 \left(\frac\varepsilon
{b^2 {\cal L}^2 \kappa^3} \right)^{1/5} ,
\label{Lkwrat}
\end{split}\end{align}
where we introduced
\begin{equation}
b = \km\ell \sim 1\ .
\end{equation}

We note that the numerical value of the prefactor in Eqs.~(\ref{Asol}) and (\ref{Lkwrat}) should be taken with caution. In the calculations we assume that the total energy of Kelvin waves can be found by the integral~(\ref{Ekw}) limited from below by $\km$ with the scale-invariant spectrum~(\ref{LN}). In reality this spectrum was derived for $k\gg \km$ while the main contribution to $E\kw$ is coming from the region $k\simeq \km$. Behavior of the Kelvin-wave spectrum in this long-wavelengths region may be different and, in general, is not universal. 

In some applications, the tilt $\theta$ of a vortex carrying Kelvin waves with respect to the direction of the straight vortex is of interest. The averaged tilt angle can be connected to the length increase
 \begin{align}\begin{split}\label{Lkw}
L_{\rm kw} &=  \int _0^L \sqrt{1+ \tan^2 \theta(z) } \, \mbox{d}z - L \\ &\simeq \frac12 \int _0^L \tan^2\theta(z) \,  \mbox{d}z\ = \frac 12 \langle \tan^2 \theta(z)  \rangle  L \ .
 \end{split} \end{align} 
  Together with Eq.~(\ref{LkwL}) this results in
  \begin{equation} \label{tan}
  \langle \tan^2 \theta(z)  \rangle  \simeq  2\,  \frac{  L_{\rm kw}}{ L} = \Psi\,,
  \end{equation}
where $\Psi$ is given by Eq.~(\ref{psisol}).

For example, let us consider a vortex of length $L=100\,\mu$m in superfluid $^4$He ($\kappa=9.9\cdot10^{-4}\,$cm$^2$/s, $\Lambda = 17$, $\rhos = 0.14\,$g/cm$^3$). Vortex is agitated with the frequency $f_0 = 30\,$kHz which we assume to set the longest Kelvin wave length $\km = \sqrt{8\pi^2f_0/\kappa\Lambda} \approx 1.2\cdot 10^4\,$cm$^{-1}$, $\lambda_{\km} \approx 5.3\,\mu$m. If the energy flux over the Kelvin-wave cascade is $\tilde\epsilon = 10^{-7}\,$erg/s, then we find that the amplitude of the waves at the largest scale is $A_{\km} \approx 0.5\,\mu$m, increase of the vortex length $L\kw \approx 48\,\mu$m and the averaged tilt angle $\langle \theta \rangle \approx 35^\circ$. We see that even such a moderate flux, which corresponds to working against the full vortex tension $\nus$ over $\tilde\epsilon/\nus f_0 \approx 0.2\,\mu$m per period of the drive, can bring the vortex on the edge of the regime where the turbulence of Kelvin waves may still be considered as weak.

To conclude, we have found the dependence of the amplitude of the Kelvin waves, of the length increase of the vortex, and of the average vortex tilt on the energy flux carried by the Kelvin-wave cascade. The results are applicable in the regime of weak turbulence of Kelvin waves, which is uniform along the vortex. We stress that the amplitude of the Kelvin waves, generated when a vortex is mechanically agitated, does not necessary coincide with the amplitude of the motion of the agitator. Solving the problem of excitation of Kelvin waves in a realistic experimental geometry remains a task for future research.

The work has been supported by the European Research Council (ERC) under the European Union's Horizon 2020 research and innovation programme (Grant Agreement No. 694248).

\def\He#1{$^{\it #1}\!$He}

\end{document}